\documentclass[12pt]{article}

\usepackage{amsmath,amssymb,subfigure,setspace}
\usepackage[dvips]{graphicx}
\usepackage[hang]{caption}

\def\blfootnote{\xdef\@thefnmark{}\@footnotetext} 

\long\def\symbolfootnote[#1]#2{\begingroup%
\def\thefootnote{\fnsymbol{footnote}}\footnote[#1]{#2}\endgroup}

\newcommand{\f}[2]{\frac{#1}{#2}}
\newcommand{\la}{\langle}
\newcommand{\ra}{\rangle}
\newcommand{\s}{\vec{s}}
\newcommand{\D}{{\cal D}}
\newcommand{\de}{\partial}
\newcommand{\Or}{\mathbf{O}}

\author{Matteo Giordano}
\title{Critical behaviour of the $O(3)$ nonlinear sigma model with
  topological term at $\theta=\pi$ from numerical simulations}
\makeatletter
\def\titolo{\@title}
\makeatother

\newcommand{\mytitle}{
\begin{center}
  \vspace*{1.0cm}
  \begin{doublespace}
    {\Large \bf \titolo}
  \end{doublespace}
  \vspace*{0.5cm}
  {\large Vicente Azcoiti${}^a$\symbolfootnote[1]{E-mail:
      azcoiti@azcoiti.unizar.es}, Giuseppe Di
    Carlo${}^b$\symbolfootnote[2]{E-mail: giuseppe.dicarlo@lngs.infn.it}, Eduardo Follana${}^a$\symbolfootnote[3]{E-mail:
      efollana@unizar.es},\\ and Matteo Giordano${}^a$\symbolfootnote[4]{E-mail: giordano@unizar.es}
      }\\
  \vspace*{0.5cm}{\normalsize
    { 
      ${}^a$ Departamento de F\'isica Te\'orica, 
      Universidad de Zaragoza, \\
      Calle Pedro Cerbuna 12,
      E--50009 Zaragoza, Spain \\
      ${}^b$ INFN, Laboratori Nazionali del Gran Sasso, \\
      I--67010 Assergi (L'Aquila), Italy
    }
  }\\  \vspace*{1.5cm}
\end{center}
}

\begin{document}

\mytitle

\begin{abstract}\normalsize
  We investigate the critical behaviour at $\theta=\pi$ of the
  two-dimensional $O(3)$ nonlinear sigma model with topological
  term on the lattice. Our method is based on numerical simulations at
  imaginary values of $\theta$, and on scaling transformations that
  allow a controlled analytic continuation to real values of
  $\theta$. Our results are compatible with a second order phase
  transition, with the critical exponent of the $SU(2)_1$
  Wess-Zumino-Novikov-Witten model, for sufficiently small values of
  the coupling. 
\end{abstract}

\newpage

\section{Introduction}
\label{sec:intro}

Quantum field theories with a topological term (``$\theta$-term'') in
the action have proved to be particularly challenging to
investigate. Such theories are related to a few important open
problems in theoretical physics, including the so-called ``strong $CP$
problem'' in strong interactions, and to interesting phenomena in
condensed matter physics, such as the quantum Hall effect (for a
recent review on theories with $\theta$-term, see Ref.~\cite{VP}). 

On the one hand, topological properties are intrinsically nonperturbative,
thus requiring a nonperturbative approach to the study of these
systems. On the other hand, the most effective of these approaches,
namely the numerical study by means of simulations in lattice field
theory, cannot be directly applied to these systems, due to the
presence of a so-called {\it sign problem}. In fact, the complex
nature of their Euclidean action prevents the computation of the
relevant functional integrals by means of the usual
importance-sampling techniques. Numerical investigations have then
required the use of techniques which allow to avoid the sign problem,
usually based on analytic continuation or on the resummation of the
contributions of the various topological sectors to the partition
function~\cite{BPW,PS,IKY,BISY,AN,AANV,ADGV0}. The basic idea of these
techniques is to modify or split the functional integral, in such a
way that the resulting expression(s) have a positive-definite
integration measure, and therefore can be treated with the usual
numerical techniques. The difficulty of dealing with an oscillatory
integrand is, however, not completely overcome, but simply shifted to
the problem of reconstructing the original functional integral, which
is usually a very delicate issue from the numerical point of view. 
It is worth noting that, beside having their own theoretical interest,
theories with a $\theta$-term share the sign problem with
finite-density QCD, and so the development of techniques and
algorithms to solve or by-pass the sign problem can have positive
consequences on the study of the QCD phase diagram by means of
numerical simulations. 

Among the various existing models, the two-dimensional $O(3)$
nonlinear sigma model with $\theta$-term ($O(3)_\theta$NL$\sigma$M)
deserves particular interest. It has been shown long ago by
Haldane~\cite{Haldane:1982rj,Haldane:1983ru} that chains of quantum
spins with antiferromagnetic interactions, in the semiclassical limit
of large but finite spin $S$, are related to this model at coupling 
$g^2=4/[S(S+1)]$, and at $\theta=0$ or $\pi$ if the spin is
respectively integer or half-integer. Haldane conjectured that quantum
spin chains for half-integer spins show a gapless spectrum, and 
correspondingly that a second-order phase transition takes place  
in the $O(3)_\theta$NL$\sigma$M at $\theta=\pi$, with vanishing of the
mass gap and recovery of parity. Arguments supporting this conjecture
have been provided in Ref.~\cite{Affleck:1991tj}. Moreover, in
Ref.~\cite{Affleck:1987ch} it has been argued that the critical theory
for generic half-integer spin antiferromagnets is the $SU(2)$ 
Wess-Zumino-Novikov-Witten (WZNW) model~\cite{WZNW1,WZNW2,WZNW3} at topological
coupling $k=1$, which in turn should determine the behaviour of the
mass gap near $\theta=\pi$. 

Numerical investigations of Haldane's conjecture have been performed,
following basically three different strategies. 
A first strategy~\cite{BPW,Bogli,dFPW} is
based on the determination of the probability distribution of the
topological charge by means of simulations at $\theta=0$, which allows
in principle to reconstruct the expectation values of the various
observables at $\theta\ne 0$. In order to achieve the very high
accuracy required by this approach, the authors of
Refs.~\cite{BPW,Bogli,dFPW} have employed a constrained
(``topological''~\cite{Bietenholz:2010xg}) 
action on a triangular lattice, which 
allows simulations by means of an efficient Wolff cluster
algorithm~\cite{Wolff:1988uh}. The parameters of the action were
chosen in order to be in the weak-coupling regime. 
Using finite size scaling theory, the authors of Ref.~\cite{BPW,Bogli,dFPW} found
a second order phase transition at $\theta=\pi$, in agreement with
Haldane's conjecture, and a finite size scaling in good agreement with
the assumption of a WZNW-type of critical behaviour. 

A second strategy~\cite{Alles:2007br} is based on the determination of
the mass gap at imaginary values of $\theta$, that can be obtained
directly by means of numerical simulations, and the subsequent
analytic continuation to real values of $\theta$, in order to check if
the mass gap vanishes at some point. The authors of
Ref.~\cite{Alles:2007br} found indeed that the mass gap 
vanishes at $\theta=\theta_c$ for some real $\theta_c$, and moreover
that $\theta_c=\pi$ within the errors, again in agreement with
Haldane's conjecture. 

Finally, the third strategy~\cite{ADGV1,ADGV2,CP1,AFV} makes use again of
numerical simulations at imaginary values of $\theta$, in order to
determine the topological charge density, and of a controlled way of
performing the analytic continuation to real $\theta$ that greatly
reduces the uncertainties connected to this process. 
Applying this strategy to the $CP^1$ model, that is expected to be
equivalent to the $O(3)$ model, the authors of Ref.~\cite{CP1} found a
richer phase structure, with a first-order phase transition at
$\theta=\pi$ for $\beta\lesssim 0.5$, and a line of second-order phase
transitions  with recovery of parity for $0.5 \lesssim \beta \lesssim
1.5$, with continuously varying critical exponent. At $\beta \simeq
1.5$ the critical exponent becomes $2$, and parity is recovered
analytically.   

In this paper we want to investigate further on this issue, by
applying the strategy of Refs.~\cite{ADGV1,ADGV2,CP1,AFV} directly to the
$O(3)_\theta$NL$\sigma$M. Our aim is to understand the origin of the
discrepancy between the results of Refs.~\cite{BPW,Bogli,dFPW} and 
those of Ref.~\cite{CP1}. Such a discrepancy could be of physical
origin, due to the actual inequivalence of the $O(3)$ and $CP^1$
models, contrary to the standard wisdom; or it could be of technical
origin, due to shortcomings of the employed strategy in dealing with
these models. 
The plan of the paper is the following. In Section \ref{sec:method} we
briefly review the method of Refs.~\cite{ADGV1,ADGV2,CP1,AFV}. In Section
\ref{sec:o3model} we describe the model of interest, discussing in
particular the theoretical 
prediction for the critical behaviour of the model at $\theta=\pi$ in
the continuum, and working out the consequences for the observables
relevant to our method. In Section \ref{sec:num_sim} we describe 
the $O(3)_\theta$NL$\sigma$M on the lattice, and we discuss the
results of our numerical simulations. Finally, Section \ref{sec:concl}
is devoted to our conclusions and to an outlook on open
problems.  Details of the numerical analysis are reported in the
Appendix.  

\section{Theories with topological term in the action and the method
  of scaling transformations} 
\label{sec:method}

In this Section we briefly describe the relevant formalism and
notation that will be used in the rest of the paper. The partition
function of a theory with a topological term in the action is of the
general form
\begin{equation}
  \label{eq:top1}
  Z(\theta) = \int \D\phi\, e^{- S[\phi] +  i\theta Q[\phi]} = e^{-VF(\theta)}\,,
\end{equation}
where $\phi$ denotes the degrees of freedom of the model, $\D\phi$
is the appropriate functional measure, $S$ is the non-topological
part of the action, and $Q$ is the quantised topological charge,
taking only integer values; moreover, $F(\theta)$ is the free energy
density and $V$ the volume of the system. Clearly, $Z(\theta)$ is a
periodic function of $\theta$, $Z(\theta+2\pi)=Z(\theta)$. In the
interesting cases, the integration measure is invariant under parity
(${\cal P}$), and $S$ is ${\cal P}$-even, while $Q$ is ${\cal
  P}$-odd. As a consequence, $Z(-\theta)=Z(\theta)$; combining this
with periodicity, we have that $Z(\pi+\theta)=Z(\pi-\theta)$. 

While at $\theta\ne 0,\pi$ parity is explicitly
broken, at $\theta=\pi$ any ${\cal P}$-odd observable has vanishing
expectation value in a finite volume; nevertheless, a phase transition
may take place at this point. A convenient order parameter is given by
the topological charge density,
\begin{equation}
  \label{eq:orpar}
\Or(\theta)\equiv  -i\f{\la Q \ra_{i\theta}}{V} = -\f{1}{V}\f{\de\log
    Z(\theta)}{\de\theta} =\f{\de F(\theta)}{\de\theta}\,,
\end{equation}
where we have introduced the notation
\begin{equation}
  \label{eq:notation}
  \la {\cal O}[\phi] \ra_{i\theta} = Z(\theta)^{-1}\int \D\phi\, e^{-
    S[\phi] +  i\theta Q[\phi]} {\cal O}[\phi]\,, 
\end{equation}
for the expectation value of the observable ${\cal O}[\phi]$. 
In the limit of infinite volume, a nonzero value of $\Or(\theta=\pi)$
indicates a first-order phase transition, while a divergent
susceptibility $\Or'(\theta=\pi)$ 
indicates a second-order phase transition, and so on.  

In order to reconstruct the behaviour of the order parameter near
$\theta=\pi$ using numerical simulations, one has to start from
imaginary values of the vacuum angle $\theta=-ih$, with $h\in
\mathbb{R}$. 
It has been suggested in Ref.~\cite{ADGV1} that a convenient observable is the
quantity 
\begin{equation}
  \label{eq:method1}
  y(z)=\f{\la Q \ra_h}{V\tanh\f{h}{2}}\,,\qquad
  z=\cosh\f{h}{2}\,,\quad z\ge 1\,.
\end{equation}
It is immediate to see that under analytic continuation $h\to i\theta$
one has
\begin{equation}
  \label{eq:method1bis}
  y(z)=-i\f{\la Q
    \ra_{i\theta}}{V\tan\f{\theta}{2}}=\f{\Or(\theta)}{\tan\f{\theta}{2}}\,,
  \qquad  z=\cos\f{\theta}{2}\,,\quad z\le 1\,.
\end{equation}
i.e., in terms of $z$ the analytic continuation is simply an
extrapolation from $z\ge 1$ to $z\le 1$. Notice that $y(1)=\f{2\la Q^2
  \ra_h}{V}$, and $y(0)=0$, with $z=0$ corresponding to
$\theta=\pi$.\footnote{One can have $y(0)\ne 0$ only if the
  topological charge density diverges at $\theta=\pi$, which seems
  unlikely.}  

The use of this observable is
suggested by the antiferromagnetic one-dimensional Ising model, where
the role of the $\theta$-term is played by the coupling with an
external imaginary magnetic field (for an even number of sites). This
model is exactly solvable, and one finds that $y$ actually depends
only on a specific combination of $z$ and of the antiferromagnetic
coupling $F$, namely $y(z,F)=Y((e^{-4F}-1)^{-\f{1}{2}}z)$. 
Although this property is exclusive of the one-dimensional Ising
model, nevertheless one can expect that a similar smooth relation
exists between $y(z)$ and $y_\lambda(z)\equiv y(e^{\f{\lambda}{2}}z)$
also in other models with $\theta$-term. The assumption usually made
is that $y(z)$ is a monotonically increasing function of $z$,
vanishing only for $z=0$ (i.e., the order parameter does not vanish
for $\theta\in(0,\pi)$); this is indeed the case for the models where
the exact solution is known. The quantity $y_\lambda$ is then a
monotonic function $y_\lambda(y)$ of $y$, with the property that 
$y_\lambda=0$ at $y=0$, so that starting from the smallest values of
$y$ that can be obtained by numerical simulations at real $h$, one can 
therefore reliably extrapolate towards $y=y_\lambda=0$, i.e., in the
region corresponding to real $\theta=-ih$. This is the advantage of
this method, based on scaling transformations, over other approaches
that involve an uncontrolled analytic continuation from imaginary
values of $\theta$. Having reconstructed $y_\lambda(y)$ in this
region, one can then easily reconstruct the order parameter at real
$\theta$. Clearly, the closer one gets to $y=0$, the better the
extrapolation is expected to be: this method is then expected to work
well in situations where the density of topological objects is small,
such as asymptotically free models at weak coupling.  

If one is interested only in the critical behaviour at $\theta=\pi$,
it is possible to determine the critical exponent without explicitly 
reconstructing the order parameter. Consider the effective exponent
$\gamma_\lambda(y)$, 
\begin{equation}
  \label{eq:gammal}
  \gamma_\lambda(y) \equiv \f{2}{\lambda}\log\f{y_\lambda(y)}{y}\,.
\end{equation}
Assuming a critical behaviour $\Or \propto z^\epsilon$ near $z=0$,
i.e., $\Or \propto (\pi-\theta)^\epsilon$ near $\theta=\pi$, 
one immediately sees that $y \propto z^{\epsilon+1}$ near $z=0$, and so
\begin{equation}
  \label{eq:critexp}
\gamma \equiv   \lim_{y\to 0} \gamma_\lambda(y) = \f{2}{\lambda}\lim_{z\to 0}
  \log\f{e^{(1+\epsilon)\f{\lambda}{2}}z^{1+\epsilon}}{z^{1+\epsilon}}  
  = 1+\epsilon\,.
\end{equation}
Analogously, assuming that $y_\lambda(y)$ is analytic at $y=0$, one
can obtain $\gamma$ from the relation $\gamma = \f{2}{\lambda}
\log\big(\f{d y_\lambda}{dy}\big|_{y=0}\big)$. 

The method outlined above has been checked against explicitly
solvable models, and successfully applied to models where the exact
solution is not known (see Refs.~\cite{ADGV1,ADGV2,CP1,AFV}). One
implicit assumption of this method is that the function $y_\lambda(y)$
has a ``reasonable'' behaviour near $y=0$, i.e., it can be well
approximated by polynomials, or ratios of polynomials, or other
``simple'' functions. If this is the case, the critical exponent can
then be obtained with fair accuracy. What has not been done yet is the
evaluation of the impact of logarithmic corrections on the reliability
of the extrapolation. The result Eq.~\eqref{eq:critexp} holds
independently of logarithmic  corrections to the critical behaviour,
i.e., it holds even if $\Or \propto z^\epsilon\log(1/z)^{-\beta}$;
nevertheless, the way in which the limit is approached in this case
can make the extrapolation more difficult. This issue will be  
discussed further on in the next Section.

\section{The $O(3)$ nonlinear sigma model with a topological 
term} 
\label{sec:o3model}

In this Section we briefly recall the main properties of the  $O(3)$
nonlinear sigma model with a topological term
($O(3)_\theta$NL$\sigma$M) in two dimensions, and we work out the
consequences of the expected critical behaviour at $\theta=\pi$ for
the method of scaling transformations described in the previous
Section.

\subsection{Critical behaviour at $\theta=\pi$}
\label{sec:crit0}

The degrees of freedom of the $O(3)_\theta$NL$\sigma$M in two
dimensions are real three-com\-po\-nent spin variables $\s(x)$ of modulus
one, $\s(x)^2=1$, ``living'' at the point $x\in
\mathbb{R}^2$. Expectation values are defined in terms of functional 
integrals as follows, 
\begin{equation}
  \label{eq:model}
  \begin{aligned}
    \la {\cal O}[\s] \ra_{i\theta} &\equiv  Z(\theta)^{-1}\int \D\s\,
    e^{- S[\s] + 
      i\theta Q[\s]}\,{\cal O}[\s]\,, \quad 
  Z(\theta) &= \int \D\s\, e^{- S[\s] +  i\theta Q[\s]}\,,
  \end{aligned}
\end{equation}
where the measure is given by $\D\s = \prod_x d^3\s(x)
\delta(1-\s(x)^2)$. In the continuum,
\begin{equation}
  \label{eq:action0}
S[\s]=  \f{1}{2g^2}\int d^2x\, \de_\mu \s(x) \cdot\de_\mu \s(x)\,,
\end{equation}
and the topological charge $Q[\s]$ is given by
\begin{equation}
  \label{eq:charge}
  Q[\s] = \f{1}{8\pi}\int d^2x\, \s(x)\cdot
     \epsilon^{\mu\nu}\de_\mu\s(x) \wedge \de_\nu\s(x)\,.
\end{equation}
Here $\mu,\nu=1,2$, and sum over repeated indices is understood; the
antisymmetric symbol $\epsilon^{\mu\nu}$ is defined as
$\epsilon^{12}=-\epsilon^{21}=1$, $\epsilon^{11}=\epsilon^{22}=0$.   

While the theory possesses a mass gap at $\theta=0$, it has been
argued that the mass gap $m(\theta)$ vanishes as $\theta\to\pi$ with 
the following behaviour~\cite{AGSZ}: 
\begin{equation}
  \label{eq:massgap}
  m(\theta) \propto |\pi-\theta|^{\f{2}{3}} \left\vert \log\f{1}{|\pi-\theta|}
  \right\vert^{-\f{1}{2}} \mathop =_{\theta<\pi,\,\theta\simeq\pi}
  (\pi-\theta)^{\f{2}{3}} \left( \log\f{1}{\pi-\theta} 
  \right)^{-\f{1}{2}}\,,
\end{equation}
where we have neglected subleading terms.\footnote{From now on we will
work in the interval $\theta\in[0,\pi]$, so that we can discard the
absolute values.} This prediction follows from
the following considerations for the continuum theory (see
Refs.~\cite{Affleck:1987ch,AGSZ,CM}). Near
$\theta=\pi$, the effective action for the $O(3)$ sigma model is given
by the $SU(2)_1$ Wess-Zumino-Novikov-Witten (WZNW)
model~\cite{WZNW1,WZNW2,WZNW3}, with a marginally irrelevant,
parity-preserving perturbation, and a relevant, parity-breaking
perturbation, whose coupling $\tilde g$ is a function 
of $(\pi-\theta)$ that vanishes at $\theta=\pi$. 
Renormalisation-group
arguments relate as follows the coupling and the correlation length
$\xi$ of the system~\cite{AGSZ}, 
\begin{equation}
  \label{eq:RG1}
  \f{1}{\tilde g} \propto \xi^{\f{3}{2}}(\log\xi)^{-\f{3}{4}}\times
  \left[1+{\cal O}\left((\log\xi)^{-1}\right)\right]\,;
\end{equation}
neglecting subleading terms, as $m\propto\xi^{-1}$, one finds
\begin{equation}
  \label{eq:RG2}
  m \propto \tilde g^{\f{2}{3}} \left(\log\f{1}{\tilde
      g}\right)^{-\f{1}{2}}\,. 
\end{equation}
It is usually assumed that $\tilde g \propto (\pi-\theta) + \ldots$, 
so that Eq.~\eqref{eq:massgap} immediately follows.

Following Kadanoff, one expects that near the critical point
$\theta=\pi$ the free energy density $F(\theta)$ 
is proportional to the square of the inverse of the 
correlation length, that in turn is proportional to the inverse mass
gap, so that
\begin{equation}
  \label{eq:fed}
  F(\theta) \propto \f{1}{\xi(\theta)^2} \propto m(\theta)^2\,.
\end{equation}
The order parameter for parity breaking $\Or(\theta)$, defined in
Eq.~\eqref{eq:orpar}, is therefore expected to show the following
behaviour near $\theta=\pi$, 
\begin{equation}
  \label{eq:orpar2}
  \Or(\theta) \propto \f{\de m(\theta)^2}{\de\theta} \propto
  (\pi-\theta)^{\f{1}{3}}\left(\log\f{1}{\pi-\theta}\right)^{-1}\,,
\end{equation}
where we have neglected subleading terms. This behaviour is 
conveniently rewritten as follows in terms of the variable
$z=\cos\f{\theta}{2}$ ($z\le 1$),
\begin{equation}
  \label{eq:orpar3}
  \Or(\theta) \propto z^{\f{1}{3}}\left(\log\f{1}{z}\right)^{-1}\,,
  \quad z\ll 1\,.
\end{equation}
The critical behaviour is therefore a second-order phase transition,
with recovery of parity, with a critical exponent $\epsilon=\f{1}{3}$.

For future utility, it is useful to work out the first correction to
the leading behaviour Eq.~\eqref{eq:RG2}. This does not require the
knowledge of the ${\cal O}((\log \xi)^{-1})$ terms in
Eq.~\eqref{eq:RG1}; we have
\begin{equation}
  \label{eq:RG2bis}
  m \propto \tilde g^{\f{2}{3}} \left(\log\f{1}{\tilde
      g}\right)^{-\f{1}{2}} \left[1- \f{3}{8}
    \f{\log\log\f{1}{\tilde g}}{\log\f{1}{\tilde g}} + 
      \f{1}{\log\f{1}{\tilde g}}r\left(\log\textstyle\f{1}{\tilde g}\right)
\right]\,,
\end{equation}
where the function
$r(x)$ is of the form $r(x)= r_0 + (r_1\log(x)+r_2)/x+\ldots$, 
and we have omitted subleading terms at large $x$. We will assume that 
subleading terms in the relation $\tilde g \propto (\pi-\theta) +
\ldots$ are suppressed as powers of $\pi-\theta$, so 
that they can be safely ignored in the analysis of the following
subsections. Using Eq.~\eqref{eq:RG2bis}, we find for
the order parameter 
\begin{equation}
  \label{eq:RG3}
  \Or   \propto
z^{\f{1}{3}} 
  \left(\log\f{1}{z}\right)^{-1}\left[1-
    \f{3}{4}\f{\log\log\f{1}{z}}{\log\f{1}{z}} + 
      \f{1}{\log\f{1}{z}}\tilde
      r\left(\log\textstyle\f{1}{z}\right) 
\right]\,,
\end{equation}
where again $\tilde r(x)$ is of the form $\tilde r(x)= \tilde r_0 +
(\tilde r_1\log(x)+ \tilde r_2)/x+\ldots$.

\subsection{Effect of logarithmic corrections on the effective
  exponent. Extension of the method} 
\label{sec:crit}

\begin{figure}[t]
  \centering
  \subfigure[]{\includegraphics[width=0.49\textwidth]{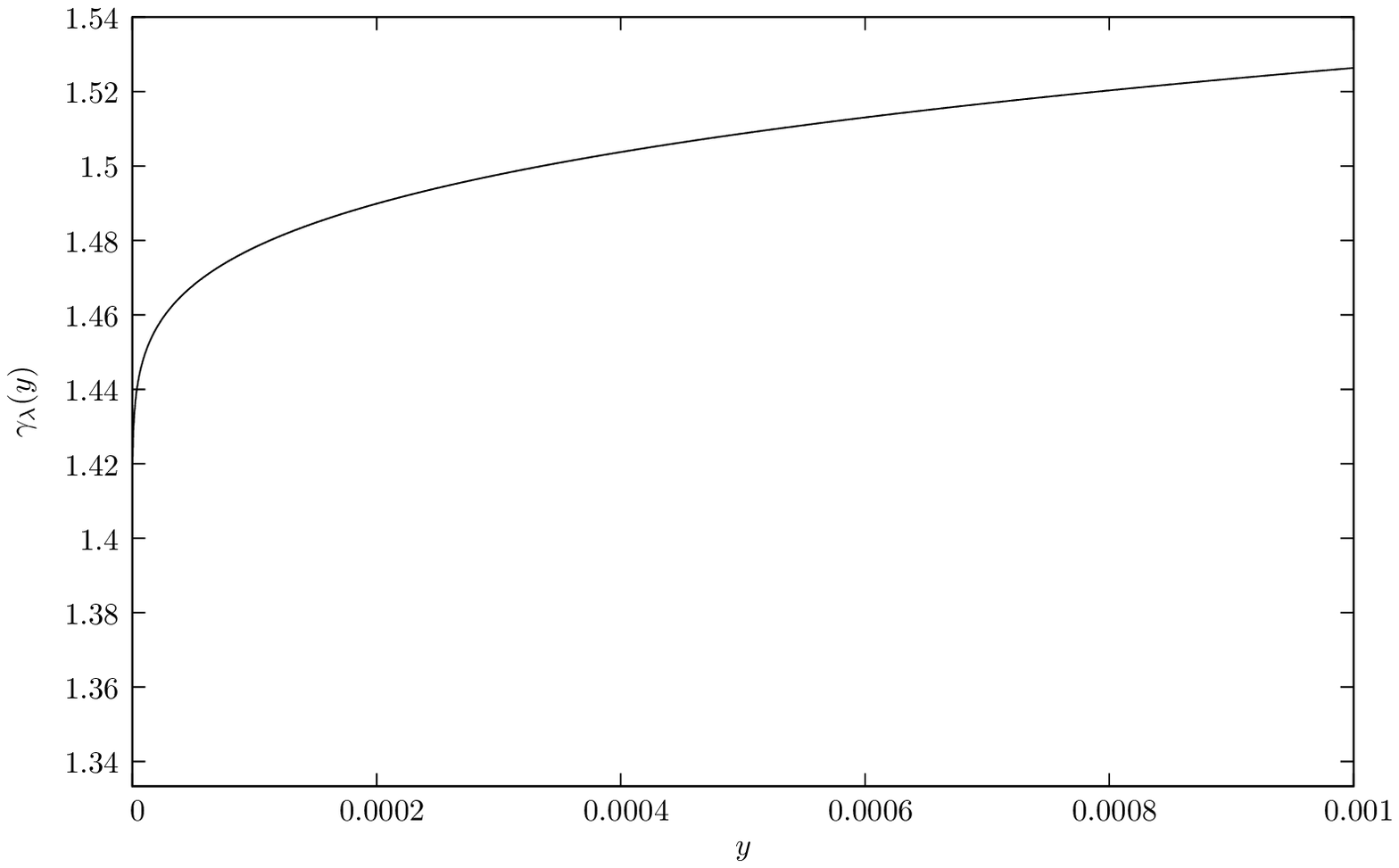}}
  \subfigure[]{\includegraphics[width=0.49\textwidth]{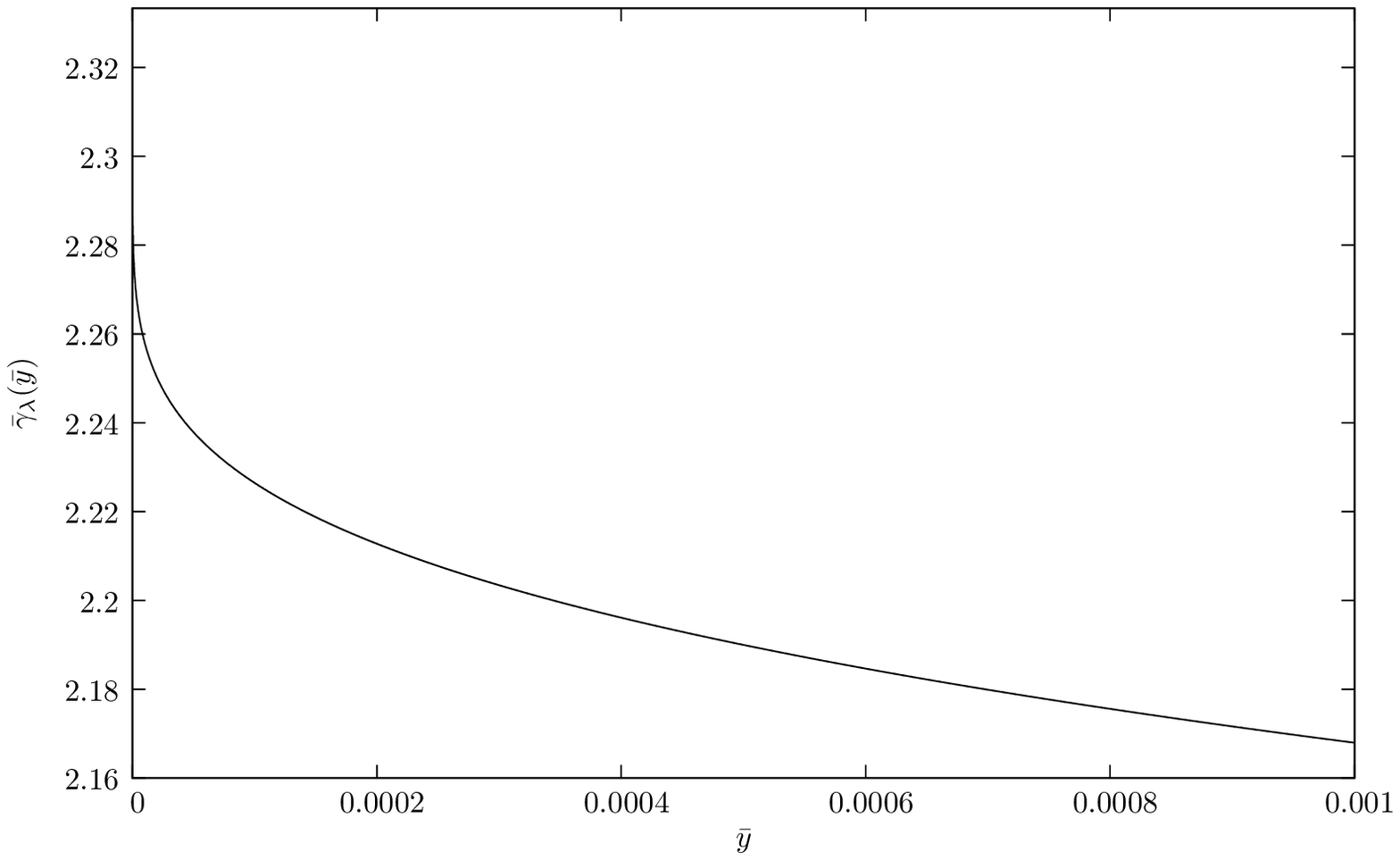}}
  \caption{(a) Theoretical prediction for $\gamma_\lambda(y)$ up to order
     ${\cal O}\left(\log\f{1}{y(z)}\right)$. \\ (b) Theoretical prediction
     for $\bar\gamma_\lambda(\bar y)$ up to order 
    ${\cal O}\left(\f{\log\log\f{1}{\bar y(z)}}{(\log\f{1}{\bar
          y(z)})^2}\right)$.}  
  \label{fig:thpred}
\end{figure}

We can now work out how the predicted behaviour of the order parameter
near the critical point reflects on the effective exponent
$\gamma_\lambda$ defined in Eq.~\eqref{eq:gammal}. Using
Eq.~\eqref{eq:RG3}, one finds that near $z=0$ the quantity $y(z)$ has the
following behaviour: 
\begin{equation}
  \label{eq:log1}
  y(z) = y_0\, z^{\f{4}{3}} \left(\log\f{1}{z}\right)^{-1}c(z)\,,
\end{equation}
where $y_0$ is some constant and $c(z)=1-
\f{3}{4}\f{\log\log\f{1}{z}}{\log\f{1}{z}} + \ldots$ (see
Eq.~\eqref{eq:RG3}), where the dots stand for subleading terms. We
have therefore for $y_\lambda(z)$
\begin{equation}
  \label{eq:log2}
  \begin{aligned}
    y_\lambda(z)&= e^{\f{\lambda}{2}\f{4}{3}}y_0\, z^{\f{4}{3}}
  \left(\log\f{1}{z}-\f{\lambda}{2}\right)^{-1}c(e^{\f{\lambda}{2}}z)\\
 &= e^{\f{\lambda}{2}\f{4}{3}}y_0\, z^{\f{4}{3}}
  \left(\log\f{1}{z}\right)^{-1}\left(1+
    \f{\lambda}{2}\f{1}{\log\f{1}{z}} +\ldots\right)c(e^{\f{\lambda}{2}}z) \\
&
=e^{\f{\lambda}{2}\f{4}{3}}\left(1+
    \f{\lambda}{2}\f{1}{\log\f{1}{z}}+\ldots\right)y(z) \,,
  \end{aligned}
\end{equation}
and so
\begin{equation}
  \label{eq:log4}
  \gamma_\lambda(y)= \f{4}{3} + \f{2}{\lambda}\log\left(1+
   \f{\lambda}{2}\f{1}{\log\f{1}{z}}\right)+\ldots
= \f{4}{3} +\f{1}{\log\f{1}{z}}+\ldots\,.
\end{equation}
Taking the logarithm on both sides of Eq.~\eqref{eq:log1}, we find
to leading order $\log\f{1}{z}=\f{3}{4}\log\f{1}{y(z)}+\ldots$, 
and plugging this into Eq.~\eqref{eq:log4} we finally obtain
\begin{equation}
  \label{eq:log7}
  \gamma_\lambda(y)= 
 \f{4}{3}\left(1
   +\f{1}{\log\f{1}{y}}\right)+o\left(\f{1}{\log\f{1}{y}}\right)\,. 
\end{equation}
The derivative of this function is infinite at the origin: as a 
consequence, the effective exponent changes abruptly at very small
$y$, going from $\gamma_\lambda= \f{4}{3}\simeq 1.33$ at $y=0$ to
$\gamma_\lambda\simeq 1.43$ at $y=10^{-6}$ (see
Fig.~\ref{fig:thpred}). From a practical point of 
view, this behaviour makes it very hard to obtain the correct
extrapolation from numerical data: one would need high precision data
at very small values of $y$ in order to figure out the logarithmic
behaviour. 

It is possible to modify the method of scaling transformations
discussed above, in order to reduce the effect of the logarithmic
corrections. Indeed, it suffices to consider a new function $\bar
y(z)$, obtained by multiplying $y$ by an appropriate factor, designed
to cancel the 
logarithmic corrections at $\theta=\pi$. A convenient choice is 
\begin{equation}
  \label{eq:extmet4}
    \bar y(z)\equiv 
    y(z)\,\cosh\f{h}{2}\,\log\left(1+\f{1}{\cosh\f{h}{2}}\right) 
    = 
      \f{\la Q 
      \ra_h}{V\tanh\f{h}{2}}
      \,\cosh\f{h}{2}\,\log\left(1+\f{1}{\cosh\f{h}{2}}
    \right)
    \,. 
\end{equation}
It is easy to show that the extra term behaves as
\begin{equation}
  \label{eq:extmet2}
 \cosh\f{h}{2} \log\left(1+\f{1}{\cosh\f{h}{2}}\right)
\mathop\to_{h\to i\theta}  
\cos\f{\theta}{2}\log\left(1+\f{1}{\cos\f{\theta}{2}}\right)
\mathop\to_{\theta\to\pi} z\log\f{1}{z}\,;
\end{equation}
therefore, near $z=0$ we have that $\bar y(z) \propto z^{\f{7}{3}}$,
without logarithmic corrections.\footnote{\label{foot:logs} More 
generally, if the order parameter behaves as $\Or
\propto z^{\epsilon}\left(\log\f{1}{z}\right)^{-\rho}$, one can 
define 
$$\bar y(z,\rho)=\f{\la Q
      \ra_h}{V\tanh\f{h}{2}}\left[\cosh\f{h}{2}\,
      \log\left(1+\f{1}{\cosh\f{h}{2}} 
    \right)\right]^\rho
    \,. $$
in order to take care of logarithmic factors.
One has that $\bar
y(z,\rho)\propto z^{\epsilon + 1 + \rho}$ near $z=0$, without
logarithmic corrections. }
Compared to similar functions yielding the desired logarithmic term,
this choice has the advantage that the behaviour
Eq.~\eqref{eq:extmet2} of the extra factor has only corrections of
order ${\cal O}(z^2)$ near $z=0$, and that the large-$h$ behaviour of
$\bar y$ is the same as that of the topological charge density, so
avoiding possible distortions in the numerical analysis. Moreover, the
extra factor is a monotonically increasing function of $z$, so that it
cannot modify the monotonicity properties of $y(z)$ (which is assumed
to be a monotonically increasing function for the whole method to work).

It is straightforward now to work out the theoretical prediction for
the behaviour of the new effective exponent
\begin{equation}
  \label{eq:new_eff_exp}
  \bar\gamma_\lambda(\bar y)\equiv \f{2}{\lambda}\log\f{\bar
    y_\lambda}{\bar y}\,. 
\end{equation}
Clearly,
\begin{equation}
  \label{eq:new_eff_exp2}
 \bar\gamma \equiv  \lim_{\bar y\to 0} \bar\gamma_\lambda(\bar y) = \gamma+1=
  \epsilon+2 = \f{7}{3}\,. 
\end{equation}
Near $z=0$, we have that (see Eq.~\eqref{eq:RG3})
\begin{equation}
  \label{eq:log1bis}
\bar  y(z) = \bar y_0\, z^{\f{7}{3}}\left[1-
    \f{3}{4}\f{\log\log\f{1}{z}}{\log\f{1}{z}} + \ldots
\right]\,, 
\end{equation}
where $\bar y_0$ is some constant, and also
\begin{equation}
  \label{eq:log2bis}
  \begin{aligned}
    \bar y_\lambda(z)&= e^{\f{\lambda}{2}\f{7}{3}}\bar y_0\, z^{\f{7}{3}}
    \left[1-
    \f{3}{4}\f{\log(\log\f{1}{z} -\f{\lambda}{2})
    }{\log\f{1}{z}-\f{\lambda}{2}} +  \ldots \right]\\
 &= e^{\f{\lambda}{2}\f{7}{3}}\bar y_0\, z^{\f{7}{3}}
\left[1-
  \f{3}{4}\left(\f{\log\log\f{1}{z}}{\log\f{1}{z}}+
  \f{\lambda}{2}\f{\log\log\f{1}{z}}{\log\f{1}{z}} +\ldots
\right)+\ldots \right]\\
&= e^{\f{\lambda}{2}\f{7}{3}}\left[1-
  \f{3}{4}\f{\lambda}{2}\f{\log\log\f{1}{z}}{(\log\f{1}{z})^2} +\ldots
\right]\bar y(z)\,, 
  \end{aligned}
\end{equation}
where the neglected terms in the last passage are of order ${\cal
  O}([\log(1/z)]^{-2})$. Taking logarithms on both sides of
Eq.~\eqref{eq:log1bis} one immediately sees that to leading order
$\log\f{1}{z} = \f{3}{7}\log\f{1}{\bar y}+\ldots$, and so one finds
that\footnote{This 
  result holds with a milder assumption on the relation between
  $\tilde g$ and $\pi -\theta$, namely that $\tilde g\propto \pi
  -\theta +\ldots$ with subleading terms suppressed with
respect to $\f{\log\log\f{1}{\pi-\theta}}{\log\f{1}{\pi-\theta}}$.
}
\begin{equation}
  \label{eq:log3bis}
\bar\gamma_\lambda (\bar y)= 
\f{7}{3} - \f{49}{12}\,\f{\log\log\f{1}{\bar
    y}}{(\log\f{1}{\bar y})^2}+\ldots\,.
\end{equation}
Although there still are logarithmic effects in the approach to the
limit value, the ``jump'' of the function between $\bar y=0$ and $\bar
y=10^{-6}$ is half as much as that of $\gamma_\lambda$ predicted above
(see Fig.~\ref{fig:thpred}). Moreover, it is easy to see that
corrections to the leading-order relation between $\log\f{1}{z}$ and
$\log\f{1}{\bar y}$ are vanishing as $\bar y\to 0$, i.e.,
$\log\f{1}{z} = \f{3}{7}\log\f{1}{\bar y}+o(1)$, while in the
relation between $\log\f{1}{z}$ and $\log\f{1}{y}$ there are also 
subleading but divergent terms as $y\to 0$. The bottom line is that
the use of $\bar\gamma_\lambda$ and $\bar y$ instead of
$\gamma_\lambda$ and $y$ is expected to improve the numerical
analysis. 

In concluding this Section, we want to add a few remarks.
First of all, we want to stress that the results of this Section are expected
to hold in the continuum limit, and they are based on the fundamental
assumption that the critical theory at $\theta = \pi$ is the $SU(2)$ WZNW model
at topological coupling $k = 1$. Furthermore, the derivation above is correct
provided that the free energy is properly renormalised. Indeed, it is known
that the topological susceptibility, as well as the higher moments of the
topological charge distribution, diverge in the continuum limit
of the $O(3)$ nonlinear sigma model (at $\theta = 0 $)
\cite{Luscher,BDSL,Nogradi}.  
As it has been suggested in~\cite{BDSL} and recently confirmed
in~\cite{Nogradi} by 
means of numerical simulations, these divergencies can be traced back to the
first coefficient in the Fourier expansion of
$F(\theta)$, i.e.,
\begin{equation}
  \label{eq:fourier}
F(\theta) = \sum_{n=1}^\infty f_n [1 -
\cos(n\theta)]  \,,
\end{equation}
with 
$f_1$ divergent and $f_n$ finite for $n>1$. In Eq.~\eqref{eq:fed} one
should therefore use $F^R(\theta) = F(\theta) - f_1^{\rm DIV}[1 -
\cos(\theta)]$, with $f_1^{\rm DIV}$ the divergent part of $f_1$; the
following results therefore hold for the renormalised quantity 
$y^R =
\f{\de F^R}{\de \theta}\left(\tan\frac{\theta}{2}\right)^{-1}$,
and the related quantities $\gamma_\lambda^R$, $\bar y^R$ and
$\bar\gamma_\lambda^R$, in the continuum. 

However, Haldane's conjecture is formulated for small but finite lattice
spacing $a$, where $f_1^{\rm DIV}=f_1^{\rm DIV}(a)$ is still
finite. We are therefore 
interested in the critical behaviour of the model at $\theta = \pi$ and at
finite $a$, so that the observable of interest 
$y_L = \f{\de
F_L}{\de \theta}\left(\tan\frac{\theta}{2}\right)^{-1}$ 
(we are using now the
subscript $L$ for lattice quantities) is a well defined, finite quantity, which
is a function of $z=\cos\frac{\theta}{2}$ and $a$, $y_L=y_L(z,a)$. Separating
now the renormalised continuum contribution from the rest, we have $y_L(z,a) =
y^R(z) + f_1^{\rm DIV}(a)\sin\theta + \delta y_L(z,a)$, where $\delta y_L$
are finite corrections that vanish as $a\to 0$. It is now evident that the
divergent term does not affect the critical behaviour at $\theta = \pi$ at
finite $a$: as it is $\propto z^2$, it is subleading with respect to $y^R(z)
\propto z^{\frac{4}{3}}$, provided that the theoretical prediction holds;
obviously, the prescription used to define the divergent part is irrelevant. On
the other hand, the corrections $\delta y_L$ may change the critical behaviour
at finite lattice spacing (and are indeed expected to do so at strong coupling):
therefore, Eqs.~\eqref{eq:log7} and \eqref{eq:log3bis} will describe
the critical behaviour of the model at 
finite $a$ only if $y^R(z)$ is the leading contribution.

We notice also that the prediction for the quantity $\bar y(z)$ has
been derived using 
the behaviour of the mass gap near the critical point $\theta =
\pi$, i.e., near $z = 0$, so that it is not expected to hold for $z
\ge 1$, where 
numerical simulations are feasible. On the other hand, due to its expected
smoothness, the prediction for $\bar y_\lambda(\bar y)$ should hold more
generally in the region of small $\bar y$, that is accessible to numerical
simulations at sufficiently small values of the coupling.

\section{Numerical simulations on the lattice}
\label{sec:num_sim}

\begin{figure}[t]
  \centering
  \subfigure[]{\includegraphics[width=0.25\textwidth]{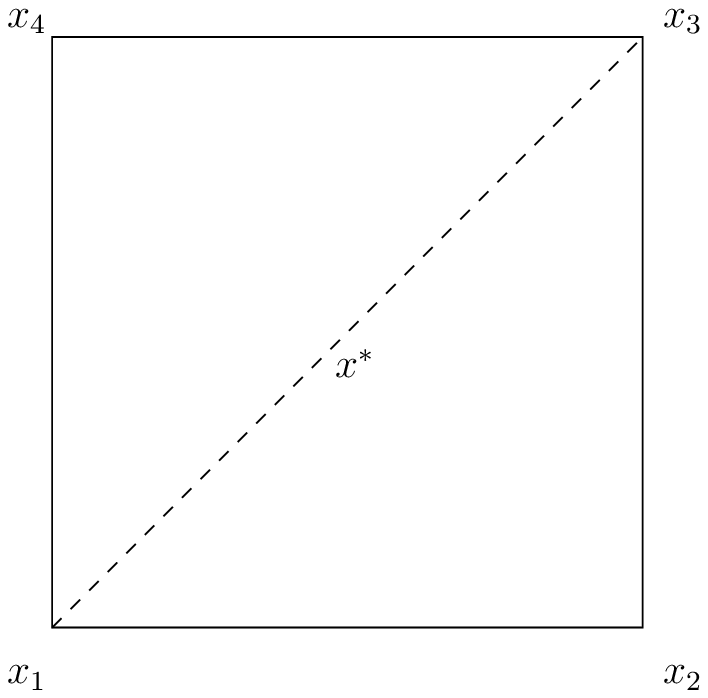}}
  \hspace{0.15\textwidth}
  \subfigure[]{\includegraphics[width=0.35\textwidth]{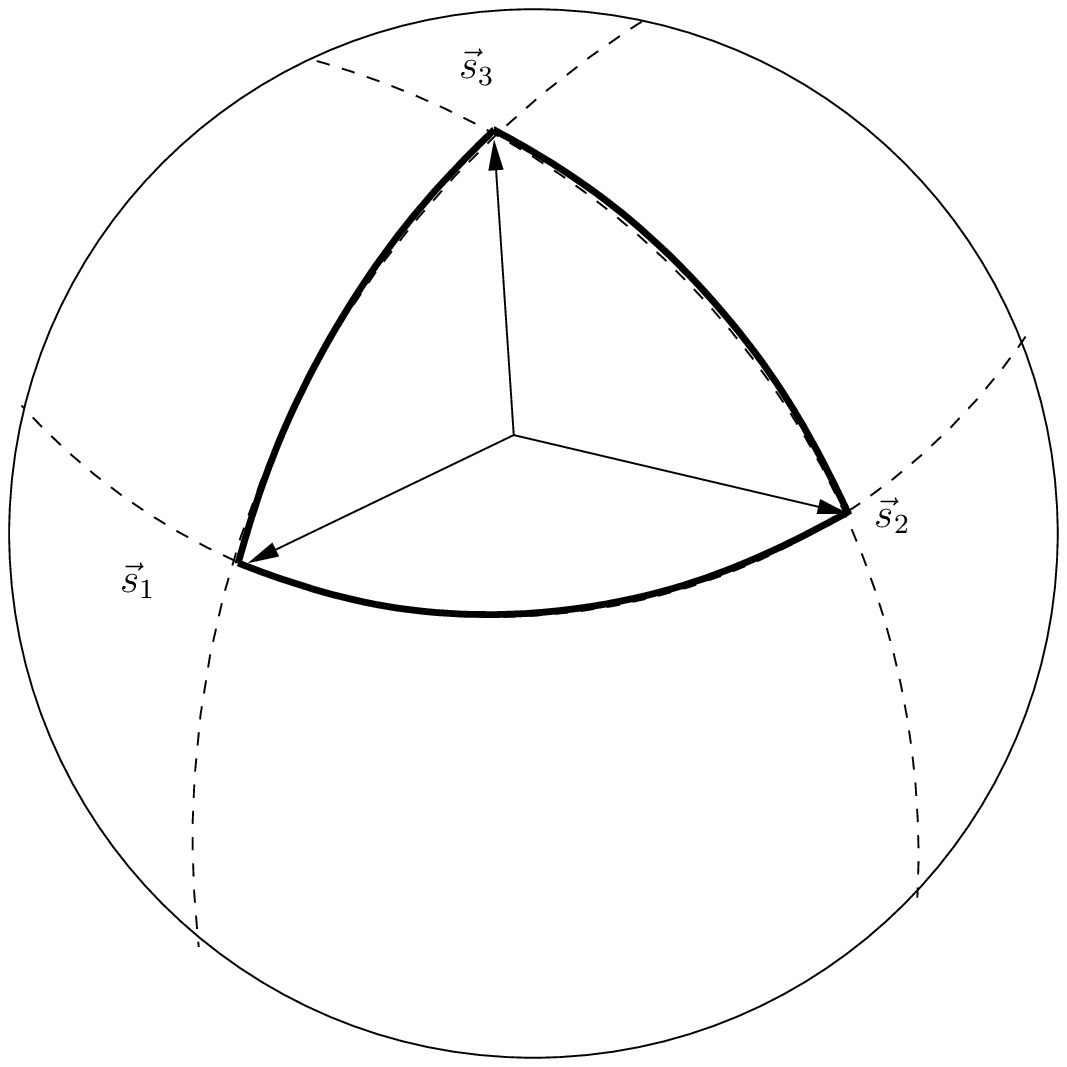}}
  \caption{(a) A unit square of the direct lattice, i.e., a site of
    the dual lattice. \\ (b) Spherical triangle corresponding to the
    spins $\s_1$, $\s_2$ and $\s_3$.} 
  \label{fig:1}
\end{figure}
In this Section we describe the setup of our numerical simulations on
the lattice, and we discuss our results on the critical behaviour of
the two-dimensional $O(3)_\theta$NL$\sigma$M at $\theta=\pi$ at finite
lattice spacing. 

\subsection{The $O(3)_\theta$NL$\sigma$M on the lattice}
\label{sec:lat_O3}

In order to compute numerically the functional integrals
Eq.~\eqref{eq:model}, one replaces the continuum by a square lattice
$\Lambda$ of finite size $V$, properly discretising the action. The
simplest choice for $S$ is  
\begin{equation}
  \label{eq:s0lat}
   S[\s] \to \frac{1}{g^2}\sum_{x\in\Lambda}\sum_{\mu=1}^2
   [1-\s(x)\cdot\s(x+\hat\mu)] = 2\beta V 
   + \beta S_{\rm latt}[\s]\,, \quad
   \beta = 1/g^2\,.
\end{equation}
Here $\hat\mu$ is a unit lattice vector in direction $\mu$. 
The lattice action $S_{\rm latt}[\s]$ is identical to the energy of
the Heisenberg statistical model, so that the resulting expression for
$Z(\theta=0)$ gives (up to an irrelevant constant) the partition
function of this model at temperature $1/\beta$ (in units of the
Boltzmann constant). As regards the topological charge, we have used
the geometrical definition of Ref.~\cite{BL}, 
\begin{equation}
  \label{eq:top_charge}
  Q_{\rm geom}[\s] = \sum_{x^*\in\Lambda^*} q(x^*)\,, \qquad
  q(x^*)=\f{1}{4\pi}\left[(\sigma A)(\s_1,\s_2,\s_3) + (\sigma
    A)(\s_1,\s_3,\s_4)\right]\,, 
\end{equation}
where $x^*$ are sites of the dual lattice $\Lambda^*$ (i.e., squares
of the direct lattice $\Lambda$), and $\s_i=\s(x_i)$ are the spin
variables living on the corners $x_i$ of the squares (ordered
counterclockwise starting from the bottom left corner, see
Fig.~\ref{fig:1} (a)). Here we have denoted by $(\sigma
A)(\s_1,\s_2,\s_3)$ the signed area of the spherical triangle having
as vertices $\s_1$, $\s_2$, and $\s_3$ (see Fig.~\ref{fig:1} (b)): the
absolute value of the area $A$ and its sign $\sigma$, i.e., the
orientation of the spherical triangle, are given respectively by  
\begin{equation}
  \label{eq:area}
 A = \alpha_1 + \alpha_2 + \alpha_3 - \pi\,, \qquad
 \sigma = {\rm sign}\left[\s_1\cdot (\s_2\wedge \s_3)\right]\,,
\end{equation}
with $\alpha_i$ the angles at the corners of the spherical
triangle; the two terms $q(x^*)=  q_1(x^*)+q_2(x^*)$ in
Eq.~\eqref{eq:top_charge} correspond to the two 
triangles in which each square on the lattice is divided. In terms of the
spin variables one has 
\begin{equation}
  \label{eq:signedarea}
  \begin{aligned}
  \exp\left\{\textstyle\f{i}{2}(\sigma A)\right\} &= \rho^{-1}\left[ 1
    + \s_1\cdot \s_2 + \s_2\cdot \s_3 + \s_3\cdot \s_1 + i \s_1\cdot
    (\s_2\wedge \s_3)\right]  \,,\\
  \rho^2 &= 2(1+\s_1\cdot \s_2)(1+\s_2\cdot \s_3)(1+\s_3\cdot \s_1)\,.
  \end{aligned}  
\end{equation}
Except for the exceptional configurations
\begin{equation}
  \label{eq:except}
  \s_1\cdot (\s_2\wedge \s_3) = 0\,, \quad 1+ \s_1\cdot \s_2 + \s_2\cdot \s_3
  + \s_3\cdot \s_1 \le 0\,,
\end{equation}
for which the topological charge is not defined, one has $\sigma=\pm
1$ and $ A < 2\pi$. One verifies directly that $Q_{\rm geom}$ has the
correct continuum limit; moreover, imposing periodic boundary
conditions, it necessarily takes only integer values.\footnote{It is
  worth mentioning that the Mermin-Wagner-Hohenberg 
  theorem~\cite{MWH1,MWH2,MWH3}, that forbids the possibility of spontaneous
  magnetisation in the model at $\theta=0$, can be easily extended to
  $\theta\ne 0$ if the geometric definition $Q_{\rm geom}$ of the
  charge is used.}  

\begin{table}[t]
  \centering
  \begin{tabular}{c|c|c}
    $\beta$ & $V$ & statistics \\ \hline
    0.9 & $100^2$ & $2\cdot 10^6$  \\
    1.2 & $100^2$ & $2\cdot 10^6$ \\
    1.5 & $100^2$ & $2\cdot 10^6$ \\
    1.6 & $200^2$ & $4\cdot 10^6$ \\
    1.7 & $350^2$ & $2\cdot 10^6$ 
  \end{tabular}
  \caption{Details of the simulations.}
  \label{tab:tech}
\end{table}
Regarding the numerical simulation of the system, the non-linear
dependence of $Q_{\rm geom}$ on the spins makes it hard to envisage fast
algorithms; we have therefore used a simple Metropolis algorithm,
supplemented by a ``partial over-relaxation'' algorithm to accelerate
the decorrelation between configurations. This ``partial
over-relaxation'' algorithm simply consists in proposing the usual
over-relaxation step used when simulating the model at $\theta=0$ to
the Metropolis accept/reject step. This algorithm turns out to be
rather efficient, especially when $\beta$ is large and the topological
content of configurations changes rarely; notwithstanding
its simplicity, it turns out also to be very effective in
decorrelating configurations.

\subsection{Numerical analysis}
\label{sec:num_an}

We have performed numerical simulations of the
$O(3)_\theta$NL$\sigma$M at various values of the coupling. For each
value of $\beta$ we have chosen 45 values of $h$, in such a way that
the topological charge was measured for both $z=\cosh\f{h}{2}$ and
$z_\lambda=e^{\f{\lambda}{2}}z\equiv\cosh\f{h_\lambda}{2}$; we used 
$\lambda=0.5$. The (real) values of $h=i\theta$ that we used lie
below the line of (possible) phase transitions determined in
Ref.~\cite{BD}, so that the region where our simulations were
performed and the real-$\theta$ axis belong to the same analyticity
domain in the complex-$\theta$ plane. The statistical error on the
topological charge has been determined through binning. The lattice
size was chosen in order for the finite size effects to be negligible
(see Tab.~\ref{tab:tech}).

\begin{figure}[t]
  \centering
  \includegraphics[width=0.7\textwidth]{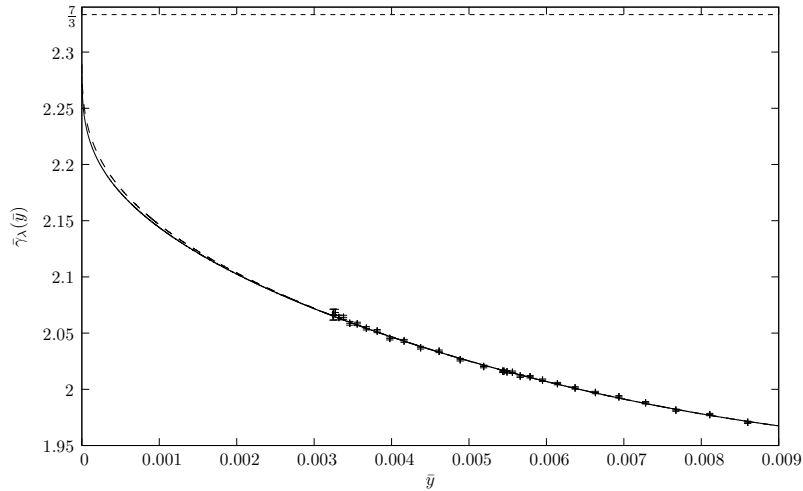}
\caption{Plot of the effective exponent $\bar\gamma_\lambda(\bar y)$
    for $\beta=1.5$, together with the result of a Bayesian fit at
    fixed $\bar\gamma=\bar\gamma_\lambda(0)$ (solid line,
    Tab.~\ref{tab:bfit1.5bis} (left)) and with free $\bar\gamma$ (long-dashed
    line, Tab.~\ref{tab:bfit1.5bis} (right)).}  
  \label{fig:fitbays1.5}
\end{figure}
We have then analysed the results for the effective exponent
$\bar\gamma_\lambda$ by means of Bayesian fits~\cite{bayes}, based on
the theoretical prediction described in Section \ref{sec:crit}. In
a nutshell, a Bayesian fit takes into account our knowledge (the
so-called {\it priors}) about the parameters that we are fitting. A
detailed account of the analysis can be found in Appendix
\ref{sec:appendix}: here we will mainly discuss the results. 

The fits were based on the following general form of
$\bar\gamma_\lambda$, 
\begin{equation}
  \label{eq:bay_zero}
  \bar\gamma_\lambda(\bar y) = \bar\gamma + F\left(\log\f{\bar
      y_0}{\bar y},\{a_j^{(k)}\}\right) \,,
\end{equation}
with $F(x,\{a_j^{(k)}\})\to 0$ as $x\to\infty$, that can be derived from the
expected critical behaviour at $\theta=\pi$ (neglecting terms that
vanish as power laws). The values of the parameters $\bar y_0$ and
$a_j^{(k)}$ are not determined by the theoretical analysis, and have
been fitted to the lattice data.

\begin{figure}[t]
  \centering
  \includegraphics[width=0.7\textwidth]{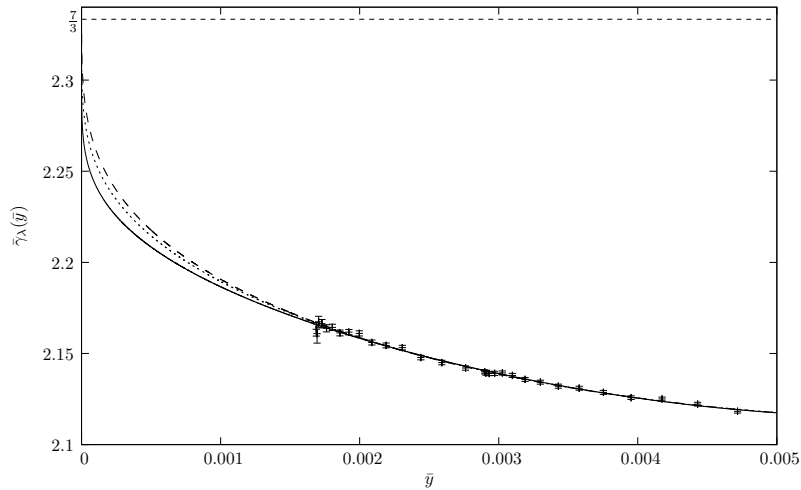}
\caption{Plot of the effective exponent $\bar\gamma_\lambda(\bar y)$
    for $\beta=1.6$, together with the result of a Bayesian fit at
    fixed $\bar\gamma=\bar\gamma_\lambda(0)$ (solid line,
    Tab.~\ref{tab:bfit1.6bis} (left)) and with free
    $\bar\gamma$ (long-dashed line,
    Tab.~\ref{tab:bfit1.6bis} (right), and short-dashed line, Tab.~\ref{tab:b3fit1.6bis}).}
  \label{fig:fitbays1.6}
\end{figure}
\begin{figure}[t]
  \centering
  \includegraphics[width=0.7\textwidth]{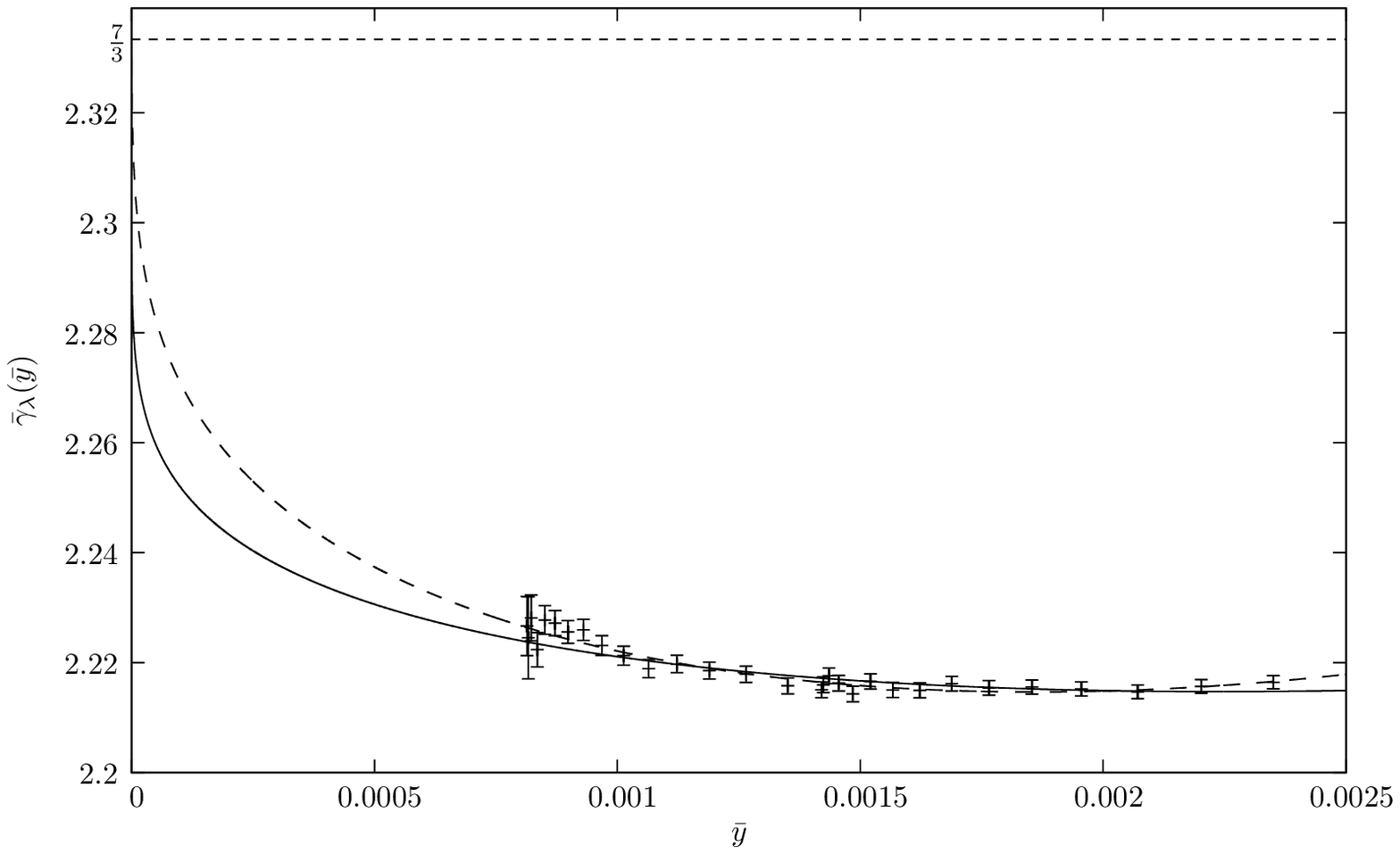}
  \caption{Plot of the effective exponent $\bar\gamma_\lambda(\bar y)$
    for $\beta=1.7$, together with the result of a Bayesian fit at
    fixed $\bar\gamma=\bar\gamma_\lambda(0)$ (solid line,
    Tab.~\ref{tab:bfit1.7bis} (left)) and with free
    $\bar\gamma$ (long-dashed line, Tab.~\ref{tab:bfit1.7bis} (right)).}
  \label{fig:fitbays1.7}
\end{figure}

A first analysis has been carried out by fixing $\bar\gamma$ to the
theoretical value, $\bar\gamma=\f{7}{3}$, and fitting the other
parameters, starting with $\bar y_0$ only and progressively adding
terms, in order of relevance. We have then used the information
obtained on $\bar y_0$ to tune the priors for a second fit, letting
all the parameters free to vary. The results are reported in
Tabs.~\ref{tab:bfit1.5bis}, \ref{tab:bfit1.6bis} and
\ref{tab:bfit1.7bis}, for $\beta=1.5$, $\beta=1.6$ and $\beta=1.7$,
respectively. Finally, at $\beta=1.6$ we have also tried a
fit using information on $a^{(1)}_0$, obtained from the fit at fixed
$\bar\gamma$, in order to set the corresponding priors: the results
are reported in Tab.~\ref{tab:b3fit1.6bis}. The results of the fit
with the largest number of parameters are shown in
Figs.~\ref{fig:fitbays1.5}, \ref{fig:fitbays1.6} and
\ref{fig:fitbays1.7}, for $\beta=1.5$, $\beta=1.6$ and $\beta=1.7$,
respectively.  

From the results of the analysis described above, we conclude that the
lattice data are compatible, within the errors, with the critical
behaviour predicted from the WZNW model, at $\beta=1.5$, $\beta=1.6$
and $\beta=1.7$. On the other hand, data at $\beta=0.9$ and
$\beta=1.2$ led to bad-quality fits when fixing $\bar\gamma$ to the
theoretical value, and to a value of $\bar\gamma$ considerably smaller
than the theoretical prediction when allowed to float ($\sim
1.9$ for $\beta=0.9$, $\sim 2$ for $\beta=1.2$). This shows that the
WZNW-like critical behaviour does not hold at small $\beta$, breaking
down at some critical value yet to be determined, but does not allow
us to draw any conclusion on the details of what happens as one lowers
$\beta$. The problem is that our analysis {\it assumes} a given
logarithmic factor in the critical behaviour of the topological charge
at $\theta=\pi$, rather than obtaining it from the numerical data. A
few attempts have shown that if we vary the exponent of the
logarithmic factor in $\bar y$, as described in footnote
\ref{foot:logs}, we still obtain a good fit but the value for the
critical exponent $\bar\gamma$ resulting from the fit changes,
too. For this reason, we have not attempted a more quantitative
analysis at $\beta=0.9$ and $\beta=1.2$.  

Summarising, our results are compatible with one of the two following
scenarios.
\begin{enumerate}
\item There is a critical value $\beta=\beta_c$, above which the
  critical behaviour of the $O(3)_\theta$NL$\sigma$M at $\theta=\pi$
  is exactly the one predicted by the WZNW model at topological
  coupling $k=1$. 
\item The critical behaviour becomes exactly the one predicted by the
  WZNW model only at infinite $\beta$, but for $\beta$ large enough,
  $\beta\gtrsim\tilde\beta_c$, the difference is not appreciable
  numerically. 
\end{enumerate}
As for what happens at small $\beta$, there are various
possibilities. As one expects the system to undergo a first-order
phase transition at $\theta=\pi$ at strong coupling, in the case of
the first scenario above there may be a sharp change in the nature of
the transition from first order to WZNW-like second order, with the
order parameter vanishing at $\theta=\pi$ as $\Or\propto
(\pi-\theta)^\epsilon$ with $\epsilon=\f{1}{3}$. It is
however also possible that the nature of the transition changes 
continuously, i.e., the critical exponent $\epsilon$ varies from 0 to
$\f{1}{3}$ as $\beta$ increases, either reaching $\f{1}{3}$ at some
finite value of $\beta=\beta_c$, or only asymptotically, as in the
second scenario above. 

\section{Conclusions}
\label{sec:concl}

In this paper we have studied the critical behaviour of the
two-dimensional $O(3)$ nonlinear sigma model with $\theta$-term
($O(3)_\theta$NL$\sigma$M) at $\theta=\pi$, by means of numerical
simulations at imaginary $\theta$. Using the method of
Refs.~\cite{ADGV1,ADGV2,CP1,AFV}, it is possible in principle to
reconstruct the behaviour of the topological charge density for real
$\theta$, and so investigate the issue of parity symmetry breaking at 
$\theta=\pi$. The theoretical expectation is that parity symmetry is
recovered at $\theta=\pi$ through a second-order phase transition,
the behaviour at the critical point being determined by the $SU(2)$ 
Wess-Zumino-Novikov-Witten (WZNW) model~\cite{WZNW1,WZNW2,WZNW3} at
topological coupling $k=1$. 

Assuming that this is the case, one can show that the method of
Refs.~\cite{ADGV1,ADGV2,CP1,AFV} is unlikely to yield the correct
critical exponent, as the large logarithmic violations to scaling at
the critical point make it difficult to reconstruct the critical
behaviour from the numerical data. Assuming that the logarithmic
violations are known, it is however easy to modify the method in order 
to overcome this problem. We have then been able to show that our
numerical results for sufficiently large $\beta$, i.e., for
sufficiently weak coupling, are compatible with
the expected WZNW-like behaviour at $\theta=\pi$, in agreement with
previous numerical investigations~\cite{BPW,Bogli,dFPW}. 

Several issues remain open. Although the modified method allows to
take care of logarithmic violations, it is necessary to know them in
advance in order for it to work properly. In fact, 
an incorrect assumption on these logarithmic violations
could not be detected from the numerical analysis, and so would lead
to an incorrect evaluation of the critical exponent. The bottom line
is that our modified method can be used to {\it test} a theoretical
expectation on the critical behaviour of a model with results from
numerical investigation, but would not lead to conclusive results if
one rather tried to {\it determine} the critical behaviour from the
numerical data.

For this reason, we have not been able to determine the critical
behaviour of the $O(3)_\theta$NL$\sigma$M at smaller values of
$\beta$, although we have been able to exclude that it is the same as
in the WZNW model. Also, due to the numerical errors, we are not able
to tell if the critical behaviour is {\it exactly} WZNW-like, starting
from some critical value of $\beta$, or if it becomes WZNW-like only
{\it asymptotically}. Further investigations are therefore required, in
order to unveil completely the phase diagram of the
$O(3)_\theta$NL$\sigma$M.  

Being in agreement with Refs.~\cite{BPW,Bogli,dFPW}, the results of this
paper are obviously in disagreement with those obtained in
Ref.~\cite{CP1} for the $CP^1$ model. There are basically two
possibilities: either the $O(3)$ and $CP^1$ model are not
equivalent, contrary to the standard wisdom; or they are equivalent,
and the results obtained in Ref.~\cite{CP1} are affected by the
numerical problems related to the logarithmic violations, discussed in 
Section~\ref{sec:crit}. In order to settle this issue, a new analysis
of the numerical data of Ref.~\cite{CP1} is required, along the lines
developed in this paper, which will be discussed in a forthcoming
publication. 

\section*{Acknowledgments}

This work was funded by an INFN-MICINN collaboration (under grant
AIC-D-2011-0663), MICINN (under grant FPA2009-09638 and
FPA2008-10732), DGIID-DGA (grant 2007-E24/2), and by the EU under
ITN-STRONGnet (PITN-GA-2009-238353). EF is supported by 
the MICINN Ramon y Cajal program. MG is supported
by MICINN under the CPAN project CSD2007-00042 from the
Consolider-Ingenio2010 program.

\newpage

\appendix

\section{Bayesian analysis}
\label{sec:appendix}

The basic idea behind constrained fits, also called Bayesian fits, is
to use the available information on a given physical problem in order
to improve the fits to the numerical data. We will not go into the details,
that can be found in Ref.~\cite{bayes} and references therein: in this
Appendix we will briefly describe the method in order to define the
relevant notation and terminology. Then, after slightly extending the
theoretical analysis of Section \ref{sec:crit}, we describe its
application to the problem at hand. 

\subsection{Constrained fits}

Constrained fits are performed in practice by minimising a modified
chi-squared, the augmented chi-squared $\chi^2_{\rm aug}$, defined
as $\chi^2_{\rm aug} = \chi^2 + \chi^2_{\rm prior}$, where $\chi^2$ is
the usual chi-squared, and where $\chi^2_{\rm prior}$ contains extra
information that is used to constrain the fit. If, thanks to prior
theoretical knowledge, one expects the parameters to be fitted, call
them $a_1,a_2,\ldots, a_n$, to be close to the values $\tilde a_1,\tilde
a_2,\ldots, \tilde a_n$ within the ranges $\tilde \sigma_1,\tilde
\sigma_2,\ldots, \tilde \sigma_n$, then one sets  
\begin{equation}
  \label{eq:priors}
  \chi^2_{\rm prior} = \sum_{i=1}^n \f{(a_i-\tilde a_i)^2}{\tilde\sigma_i^2}\,.
\end{equation}
This approach allows to add as many terms to the fitting function as
desired, contrary to what happens with standard fits. The goodness of
the procedure is judged by the convergence of the errors on the
various parameters as the number of terms is increased, and by the
value of $\f{\chi^2_{\rm aug}}{n^\circ {\rm data}}$, which should be
of the order of or smaller than 1. If this is the case, the resulting
error at the end of the procedure is expected to give a reasonable
estimate of both the statistical and the systematic errors on the
parameters of the fit.

\subsection{Subleading terms in the effective exponent}

In order to perform a Bayesian analysis of our numerical data, we need 
to go a few steps further in the derivation of the theoretical
prediction for the relevant quantities. Ignoring
corrections that contain powers of $z$, one can show that the
effective exponent $\bar  \gamma_\lambda(\bar y)$ is of the following
form, 
\begin{equation}
  \label{eq:ad1}
\bar  \gamma_\lambda(\bar y) = \f{7}{3}\left\{ 1 + \f{1}{\log\f{\bar
      y_0}{\bar y}} \sum_{k=1}^\infty\sum_{j=0}^k
  Y^{(k)}_j(\lambda)\f{\left(\log\log\f{\bar y_0}{\bar
        y}\right)^j}{\left(\log\f{\bar 
      y_0}{\bar y}\right)^k}\right\}\,.
\end{equation}
In order to determine all the coefficients $Y^{(k)}_j(\lambda)$, one
should know the subleading terms in the relation between the mass
gap and the coupling $\tilde g$, and moreover all the proportionality
constants relating $\tilde g$ with $\pi-\theta$, the free energy with
the squared mass gap, and so on. However, the coefficients
$Y^{(k)}_k(\lambda)$ are under control, as they are not affected by
the subleading terms and by the unknown constants (assuming of course
that $\tilde g$ has power, or power/log corrections only, beside the
term linear in $\pi-\theta$). Explicitly,
$Y^{(k)}_k(\lambda)=Y^{(k)}_k = (-7/4)^k$. One can therefore resum the
corresponding terms in Eq.~\eqref{eq:ad1}, obtaining 
\begin{multline}
  \label{eq:ad2}
  \bar  \gamma_\lambda(\bar y) = \f{7}{3}\left\{ 1
    -\f{7}{4}\f{\log\log\f{\bar y_0}{\bar 
        y}}{\left(\log\f{\bar 
      y_0}{\bar y}\right)^2+ \f{7}{4}\log\log\f{\bar y_0}{\bar 
        y}\log\f{\bar 
      y_0}{\bar y} }  \right. \\ \left. + \f{1}{\log\f{\bar
      y_0}{\bar y}} \sum_{k=1}^\infty\sum_{j=0}^{k-1}
  Y^{(k)}_j(\lambda)\f{\left(\log\log\f{\bar y_0}{\bar
        y}\right)^j}{\left(\log\f{\bar 
      y_0}{\bar y}\right)^k}\right\}\,.
\end{multline}
Notice that $Y^{(k)}_j(\lambda)=\left(\f{7}{3}\right)^k\tilde
Y^{(k)}_j(\lambda)$, with the coefficients $\tilde
Y^{(k)}_j(\lambda)$ expected to be of order ${\cal O}(10^{-1})$, as
they contain factors of positive powers of $\f{3}{4}$, and also powers
of $\f{\lambda}{2}$ (recall that we used $\lambda=0.5$ in our analysis).  

We have then tried a Bayesian fit retaining the first few
terms of Eq.~\eqref{eq:ad2}, namely using the following function,
\begin{multline}
  \label{eq:ad3}
   \bar  \gamma_\lambda(\bar y) = \bar \gamma\left\{ 1
    -\eta\f{\log\log\f{\bar y_0}{\bar 
        y}}{\left(\log\f{\bar 
      y_0}{\bar y}\right)^2+ \eta\log\log\f{\bar y_0}{\bar 
        y}\log\f{\bar 
      y_0}{\bar y} }  \rule{0cm}{1cm}\right.\\ \left. \rule{0cm}{1cm}
  +  \bar \gamma f_1(\bar y) + \bar \gamma^2 f_2(\bar y)  +\bar
  \gamma^3 f_3(\bar y) 
  +\bar \gamma^4 f_4(\bar y) \right\}\,.
\end{multline}
The powers of $\bar\gamma$ have been chosen so that the coefficients
in the functions $f_i$,  
\begin{equation}
  \label{eq:ad4}
  \begin{aligned}
    f_1(\bar y) &= \f{a^{(1)}_0}{\left(\log\f{\bar 
          y_0}{\bar y}\right)^2} \,, \qquad\quad
    f_2(\bar y) = \f{a^{(2)}_0 + a^{(2)}_1\log\log\f{\bar y_0}{\bar 
        y}}{\left(\log\f{\bar 
          y_0}{\bar y}\right)^3} \,,\\
    f_3(\bar y) &= \f{a^{(3)}_0+ a^{(3)}_1\log\log\f{\bar y_0}{\bar 
        y}+ a^{(3)}_2\left(\log\log\f{\bar y_0}{\bar 
        y}\right)^2}{\left(\log\f{\bar 
          y_0}{\bar y}\right)^4}\,,\\
    f_4(\bar y) &= \f{a^{(4)}_0+ a^{(4)}_1\log\log\f{\bar y_0}{\bar 
        y}+ a^{(4)}_2\left(\log\log\f{\bar y_0}{\bar 
        y}\right)^2+a^{(4)}_3\left(\log\log\f{\bar y_0}{\bar 
        y}\right)^3}{\left(\log\f{\bar 
          y_0}{\bar y}\right)^5}\,,
  \end{aligned}
\end{equation}
are at most of order $1$, and actually expected to be of order ${\cal
  O}(10^{-1})$, as explained above. Here
$\eta=\bar\gamma/(\bar\gamma-1)$: this relation is easily found by
substituting $\bar\gamma$ to the value $\f{7}{3}$, obtained for the
WZNW model, in the theoretical analysis.  

\subsection{Details of the numerical analysis}

As already explained in Section \ref{sec:num_an}, a first analysis has
been carried out by fixing $\bar\gamma$ to the theoretical value,
$\bar\gamma=\f{7}{3}$, and fitting the other parameters, starting with
$\bar y_0$ only and progressively adding terms, in order of relevance. 
As regards the priors, we assumed a Gaussian distribution for the
coefficients $a^{(k)}_j$, with central value 0 and $\sigma\approx
0.1$, while for $\bar y_0$ we have chosen central value 0 and
$\sigma\approx 2$.  
We report in Tabs.~\ref{tab:bfit1.5bis} (left), \ref{tab:bfit1.6bis} (left)
and \ref{tab:bfit1.7bis} (left) the results for $\bar y_0$ and for the
$\chi_{\rm aug}^2/{\rm n}^\circ\, {\rm data}$, for $\beta=1.5$,
$\beta=1.6$ and $\beta=1.7$, respectively. 

We have then used the information obtained on $\bar y_0$ to tune the
priors for a second fit, letting all the parameters free to vary. The
central value for $\bar y_0$ was chosen close to the result obtained
with the first fit, with $\sigma$ equal to the corresponding error.
The results for $\bar\gamma$ and $\bar y_0$ are reported in
Tabs.~\ref{tab:bfit1.5bis} (right), \ref{tab:bfit1.6bis} (right) and
\ref{tab:bfit1.7bis} (right), for $\beta=1.5$, $\beta=1.6$ and
$\beta=1.7$, respectively. At $\beta=1.6$ we have also tried a fit in
which we have used information on $a^{(1)}_0$, obtained from the fit
at fixed $\bar\gamma$, to similarly set the corresponding priors: the
results are reported in Tab.~\ref{tab:b3fit1.6bis}.

\newpage

\begin{table}[t]
  \centering
  \begin{tabular}{ccc||cccc}
    n${}^\circ$ par. & $\log\bar y_0$ & 
    $\f{\chi_{\rm aug}^2}{{\rm n}^\circ\, {\rm data}}$ & n${}^\circ$ par. & $\bar\gamma$  & $\log\bar y_0$ & 
    $\f{\chi_{\rm aug}^2}{{\rm n}^\circ\, {\rm data}}$\\ \hline
    1       &	     	   -2.3078(22) 	     &	    12   &   2 	&    2.3032(18) &   -2.051(17) 	&       2.1 \\  
    2       & 	     	   -2.699(13) 	     &	    .64  &   3 	&    2.3360(46) &   -2.726(48) 	&       .75 \\  
    3       &  	     	   -2.685(31) 	     &	    .63  &   4 	&    2.3339(87) &   -2.69(14) 	&       .75 \\  
    4       &  	     	   -2.65(20) 	     &	    .63  &   5 	&    2.3462(91) &   -2.42(15)	&       .62 \\  
    5       &  	     	   -2.55(25)	     &	    .62  &   6 	&    2.343(11)	&   -2.38(16)	&       .61 \\  
    6       &  	     	   -2.55(26)	     &	    .62  &   7 	&    2.348(12)	&   -2.36(18)	&       .60  \\ 
    7       & 	     	   -2.44(28)	     &	    .61  &   8 	&    2.346(11)	&   -2.25(18)	&       .57 \\  
    8       &  	     	   -2.39(32)	     &	    .60  &   9 	&    2.344(13)	&   -2.23(20)	&       .56 \\  
    9       &  	     	   -2.39(32)	     &	    .60  &  10 	&    2.346(15)	&   -2.24(21)	&       .56 \\  
    10      &  	     	   -2.36(34)	     &	    .59  &  11 	&    2.347(14)	&   -2.21(20)	&        .55  \\
    11      &  	     	   -2.31(34)	     &	    .59  &  12 	&    2.345(12)	&   -2.19(21)	&       .55     
  \end{tabular}
  \caption{(Left) Result of a Bayesian fit at $\beta=1.5$ with $\bar\gamma$
    fixed.\\ (Right) Result of a Bayesian fit at $\beta=1.5$ with $\bar\gamma$ free.} 
  \label{tab:bfit1.5bis}
\end{table}

\begin{table}[b]
  \centering
  \begin{tabular}{ccc||cccc}
n${}^\circ$ par. & $\log\bar y_0$ & 
    $\f{\chi_{\rm aug}^2}{{\rm n}^\circ\, {\rm data}}$ & n${}^\circ$ par. & $\bar\gamma$   & $\log\bar y_0$ &
    $\f{\chi_{\rm aug}^2}{{\rm n}^\circ\, {\rm data}}$\\ \hline 
    1       &     -1.3427(47) & 	       2.9 &       2  &     2.3222(20) &      -1.136(40) &    2.6 \\
    2       &        0.08(18) & 	       2.2 &       3  &     2.3633(61) &      -2.935(87) &   .62\\ 
    3       &        0.13(32) & 	       2.2 &       4  &     2.367(10)  &      -2.99(14)  &    .60\\ 
    4       &       -2.30(34) & 	       1.2 &       5  &     2.3711(86) &      -2.77(22) &     .55 \\
    5       &       -2.59(26) & 	       1.1 &       6  &      2.373(10) &      -2.79(18) &     .55\\ 
    6       &       -2.96(40) & 	       1.1 &       7  &      2.375(10) &      -2.76(21) &     .54\\ 
    7       &       -2.83(44) & 	       1.1 &       8  &      2.374(14) &      -2.71(28) &     .53\\ 
    8       &       -2.69(37) & 	       .99 &       9  &      2.375(14) &      -2.72(28) &     .53\\ 
    9       &       -2.80(37) & 	       .95 &       10 &      2.377(15) &      -2.73(29) &     .53\\ 
    10      &       -2.82(41) & 	       .95 &       11 &      2.377(15) &      -2.72(29) &     .53\\ 
    11      &       -2.69(36) & 	       .93 &       12 &	     2.378(17) &      -2.73(32) &     .53   
  \end{tabular}
  \caption{(Left) Result of a Bayesian fit at $\beta=1.6$ with
    $\bar\gamma$ fixed. \\ (Right) Result of a Bayesian fit at $\beta=1.6$ with $\bar\gamma$ free.}
  \label{tab:bfit1.6bis}
\end{table}
\begin{table}[t]
  \centering
  \begin{tabular}{cccc}
n${}^\circ$ par. & $\bar\gamma$  & $\log\bar y_0$ & 
    $\f{\chi_{\rm aug}^2}{{\rm n}^\circ\, {\rm data}}$\\ \hline 
  2 &	    2.3050(15) &	   -2.230(47)   &    11 \\ 
  3 &	    2.3627(61) &	   -2.927(88)   &    .58  \\ 
  4 &	    2.3597(67) &	   -2.892(91)   &    .56 \\ 
  5 &	    2.3601(71) &	   -2.87(19) 	&    .56 \\ 
  6 &	    2.3581(76) &	   -2.83(19) 	&    .55 \\ 
  7 &	    2.3586(91) &	   -2.82(21) 	&    .55 \\ 
  8 &	    2.3580(90) &	   -2.79(23) 	&    .55 \\ 
  9 &	    2.357(10)  &	   -2.77(25) 	&    .55 \\ 
 10 &	    2.357(12)  &	   -2.78(25) 	&    .55 \\ 
 11 &	    2.357(12)  &	   -2.77(26) 	&    .55 \\ 
 12 &	    2.357(13)  &	   -2.76(28) 	&    .55 
  \end{tabular}
  \caption{Result of a Bayesian fit at $\beta=1.6$ with $\bar\gamma$ free (2).}
  \label{tab:b3fit1.6bis}
\end{table}

\begin{table}[b]
  \centering
  \begin{tabular}{ccc||cccc}
    n${}^\circ$ par. & $\log\bar y_0$ & 
    $\f{\chi_{\rm aug}^2}{{\rm n}^\circ\, {\rm data}}$ & n${}^\circ$ par. & $\bar\gamma$   & $\log\bar y_0$ & 
    $\f{\chi_{\rm aug}^2}{{\rm n}^\circ\, {\rm data}}$\\ \hline
1  	 &   0.200(13)	     &    7.3    &   2 &	 2.3052(28)    	&   1.10(19)  & 4.8    \\ 
2  	 &  -2.740(52) 	     &    1.3    &   3 &	 2.3654(98)   	&  -3.35(15)  & .87     \\ 
3  	 &  -2.63(10) 	     &    1.2    &   4 &	 2.367(11)   	&  -3.35(16)  & .85   \\ 
4  	 &  -2.41(21) 	     &    1.2    &   5 &	 2.373(11)   	&  -3.07(19)  & .72    \\ 
5  	 &  -2.39(24) 	     &    1.2    &   6 &	 2.373(11)   	&  -3.08(19)  & .72     \\
6  	 &  -2.30(29) 	     &    1.2    &   7 &	 2.376(12)   	&  -2.99(20)  & .68    \\ 
7  	 &  -2.21(29) 	     &    1.2    &   8 &	 2.375(11)   	&  -2.85(21)  & .63    \\ 
8  	 &  -2.23(33) 	     &    1.2    &   9 &	 2.375(12)   	&  -2.86(21)  & .63    \\ 
9  	 &  -2.18(36) 	     &    1.2    &  10 &	 2.377(12)      &  -2.83(22)  & .62     \\ 
10 	 &  -2.14(36) 	     &    1.2    &  11 &	 2.377(12)      &  -2.77(22)  & .60     \\ 
11 	 &  -2.12(38) 	     &    1.1    &  12 &	 2.375(12)      &  -2.71(22)  & .58       
  \end{tabular}                   
  \caption{(Left) Result of a Bayesian fit at $\beta=1.7$ with
    $\bar\gamma$ fixed.\\
  (Right) Result of a Bayesian fit at $\beta=1.7$ with $\bar\gamma$ free.}
  \label{tab:bfit1.7bis}
\end{table}

\cleardoublepage


\begin{thebibliography}{99}

\bibitem{VP} E.~Vicari and H.~Panagopoulos,
  Phys.\ Rept.\  {\bf 470} (2009) 93
  [arXiv:0803.1593 [hep-th]].

\bibitem{BPW}
  W.~Bietenholz, A.~Pochinsky and U.~J.~Wiese,
  Phys.\ Rev.\ Lett.\  {\bf 75} (1995) 4524
  [hep-lat/9505019].

\bibitem{PS} J.~C.~Plefka, S.~Samuel, Phys.\ Rev.\ D {\bf 56} (1997) 44.

\bibitem{IKY} M.~Imachi, S.~Kanou, H.~Yoneyama,
  Prog.\ Theor.\ Phys.\ {\bf 102} (1999) 653.

\bibitem{BISY} R.~Burkhalter, M.~Imachi, Y.~Shinno, H.~Yoneyama, Prog.\
Theor.\ Phys.\ {\bf 106} (2001) 613.

\bibitem{AN} K.~N.~Anagnostopoulos, J.~Nishimura, Phys.\ Rev.\ D {\bf
    66} (2002) 106008.

\bibitem{AANV} J.~Ambj\o rn, K.~N.~Anagnostopoulos, J.~Nishimura,
  J.~J.~M.~Verbaarshot, JHEP {\bf 0210} (2002) 062. 

\bibitem{ADGV0} V.~Azcoiti, G.~Di Carlo, A.~Galante, V.~Laliena, Phys.\ Rev.\
 Lett.\ {\bf 89} (2002) 141601.

\bibitem{Haldane:1982rj}
  F.~D.~M.~Haldane,
  Phys.\ Lett.\ A {\bf 93} (1983) 464.

\bibitem{Haldane:1983ru}
  F.~D.~M.~Haldane,
  Phys.\ Rev.\ Lett.\  {\bf 50} (1983) 1153.

\bibitem{Affleck:1991tj}
  I.~Affleck,
  Phys.\ Rev.\ Lett.\  {\bf 66} (1991) 2429.

\bibitem{Affleck:1987ch}
  I.~Affleck and F.~D.~M.~Haldane,
  Phys.\ Rev.\ B {\bf 36} (1987) 5291.

\bibitem{WZNW1} J.~Wess and B.~Zumino, Phys.\ Lett.\ B {\bf 37} (1971)
  95.

\bibitem{WZNW2} S.~P.~Novikov, Sov.\ Math.\ Dokl.\ {\bf 24} (1981) 222.

\bibitem{WZNW3} E.~Witten, 
Commun.\ Math.\ Phys.\ {\bf 92} (1984) 455.

\bibitem{Bogli}
  M.~B\"ogli, F.~Niedermayer, M.~Pepe and U.~-J.~Wiese,
  JHEP {\bf 1204} (2012) 117
  [arXiv:1112.1873 [hep-lat]].

\bibitem{dFPW} P.~de Forcrand, M.~Pepe, and U.J.~Wiese, Phys.\ Rev.\ D
  {\bf 86} (2012) 075006 [arXiv:1204.4913 [hep-lat]].

\bibitem{Bietenholz:2010xg}
  W.~Bietenholz, U.~Gerber, M.~Pepe and U.~-J.~Wiese,
  JHEP {\bf 1012} (2010) 020
  [arXiv:1009.2146 [hep-lat]].

\bibitem{Wolff:1988uh}
  U.~Wolff,
  Phys.\ Rev.\ Lett.\  {\bf 62} (1989) 361.

\bibitem{Alles:2007br}
  B.~Alles and A.~Papa,
  Phys.\ Rev.\ D {\bf 77} (2008) 056008
  [arXiv:0711.1496 [cond-mat.stat-mech]].

\bibitem{ADGV1}
  V.~Azcoiti, G.~Di Carlo, A.~Galante and V.~Laliena,
  Phys.\ Lett.\ B {\bf 563} (2003) 117
  [hep-lat/0305005].

\bibitem{ADGV2}
  V.~Azcoiti, G.~Di Carlo, A.~Galante and V.~Laliena,
  Phys.\ Rev.\ D {\bf 69} (2004) 056006
  [hep-lat/0305022].

\bibitem{CP1}
  V.~Azcoiti, G.~Di Carlo and A.~Galante,
  Phys.\ Rev.\ Lett.\  {\bf 98} (2007) 257203
  [arXiv:0710.1507 [hep-lat]].

\bibitem{AFV}
  V.~Azcoiti, E.~Follana and A.~Vaquero,
  Nucl.\ Phys.\ B {\bf 851} (2011) 420
  [arXiv:1105.1020 [hep-lat]].

\bibitem{AGSZ} I.~Affleck, D.~Gepner, H.~J.~Schulz, and T.~Ziman, {
    J.\ Phys.\ A: Math.\ Gen.\ } {\bf 22} (1989) 511.

\bibitem{CM}
  D.~Controzzi and G.~Mussardo,
  Phys.\ Rev.\ Lett.\  {\bf 92} (2004) 021601
  [hep-th/0307143].


\bibitem{Luscher} M.~L\"uscher, Nucl. Phys. B {\bf 200} (1982) 61.

\bibitem{BDSL} G.~Bhanot, R.F.~Dashen,
  N.~Seiberg and H.~Levine, Phys. Rev. Lett. {\bf 53} (1984) 519.

\bibitem{Nogradi} D.~N\'ogr\'adi, JHEP {\bf 1205} (2012)
  089 [arXiv:1202.4616].


\bibitem{BL} B.~Berg and M.~L\"uscher, { Nucl. Phys. B} {\bf 190}
    [FS3] (1981) 412. 

\bibitem{MWH1} N.~D.~Mermin and H.~Wagner, Phys.\ Rev.\ Lett.\ {\bf
    17}, (1966) 1133.

\bibitem{MWH2} P.~C.~Hohenberg, Phys.\ Rev.\ {\bf 158} (1967) 383.

\bibitem{MWH3} S.~Coleman, Commun.\ Math.\ Phys.\ {\bf 31} (1973) 259.

\bibitem{BD} G.~Bhanot and F.~David, Nucl.\ Phys.\ B {\bf 251} [FS13]
  (1985) 127.

\bibitem{bayes} G.~P.~Lepage, B.~Clark, C.~T.~H.~Davies,
  K.~Hornbostel, P.~B.~Mackenzie, C.~Morningstar and H.~Trottier, 
  Nucl.\ Phys.\ Proc.\ Suppl.\  {\bf 106} (2002) 12
  [hep-lat/0110175].


\end{thebibliography}
\end{document}